\documentclass[aps,prb,twocolumn,superscriptaddress]{revtex4-1}
\usepackage{graphicx}
\begin{document}

%\preprint{}

\title{Highly Non-linear Excitonic Zeeman Spin-Splitting in Composition-Engineered Artificial Atoms}

\author{V. Jovanov}
\email[]{jovanov@wsi.tum.de}
\affiliation{Walter Schottky Institut and Physik Department, Technische Universit\"{a}t M\"{u}nchen, Am Coulombwall 4, 85748 Garching, Germany}
\author{T. Eissfeller}
\affiliation{Walter Schottky Institut and Physik Department, Technische Universit\"{a}t M\"{u}nchen, Am Coulombwall 4, 85748 Garching, Germany}
\author{S. Kapfinger}
\affiliation{Walter Schottky Institut and Physik Department, Technische Universit\"{a}t M\"{u}nchen, Am Coulombwall 4, 85748 Garching, Germany}
\author{E. C. Clark}
\affiliation{Walter Schottky Institut and Physik Department, Technische Universit\"{a}t M\"{u}nchen, Am Coulombwall 4, 85748 Garching, Germany}
\author{F. Klotz}
\affiliation{Walter Schottky Institut and Physik Department, Technische Universit\"{a}t M\"{u}nchen, Am Coulombwall 4, 85748 Garching, Germany}
\author{M. Bichler}
\affiliation{Walter Schottky Institut and Physik Department, Technische Universit\"{a}t M\"{u}nchen, Am Coulombwall 4, 85748 Garching, Germany}
\author{J. G. Keizer}
\affiliation{Department of Applied Physics, Eindhoven University of Technology, PO Box 513, 5600 MB Eindhoven, The Netherlands}
\author{P. M. Koenraad}
\affiliation{Department of Applied Physics, Eindhoven University of Technology, PO Box 513, 5600 MB Eindhoven, The Netherlands}
\author{M. S. Brandt}
\affiliation{Walter Schottky Institut and Physik Department, Technische Universit\"{a}t M\"{u}nchen, Am Coulombwall 4, 85748 Garching, Germany}
\author{G. Abstreiter}
\affiliation{Walter Schottky Institut and Physik Department, Technische Universit\"{a}t M\"{u}nchen, Am Coulombwall 4, 85748 Garching, Germany}
\author{J. J. Finley}
\affiliation{Walter Schottky Institut and Physik Department, Technische Universit\"{a}t M\"{u}nchen, Am Coulombwall 4, 85748 Garching, Germany}
%\thanks{}

%\date{\today}

\begin{abstract}
Non-linear Zeeman splitting of neutral excitons is observed in composition engineered In$_{x}$Ga$_{1-x}$As self-assembled quantum dots and its microscopic origin is explained. Eight-band $\textbf{k}\cdot\textbf{p}$ simulations, performed using realistic dot parameters extracted from cross-sectional scanning tunneling microscopy, reveal that a quadratic contribution to the Zeeman energy originates from a spin dependent mixing of heavy and light hole orbital states in the dot. The dilute In-composition ($x<0.35$) and large lateral size ($40-50$~nm) of the quantum dots investigated is shown to strongly enhance the non-linear excitonic Zeeman gap, providing a blueprint to enhance such magnetic non-linearities via growth engineering.
\end{abstract}

%\keywords{}

\maketitle

%----------------------------------------------------------------------------------------------------------------------------------------
% INTRODUCTION AND MOTIVATION
%
%----------------------------------------------------------------------------------------------------------------------------------------
Over the past decade, semiconductor quantum dots (QDs) have attracted significant interest, mainly due to the prospect for their use as integrated, electro-optically addressable quantum systems capable of storing and processing quantum information.\cite{Hanson2007} Quantum information processing requires the possibility for selective manipulation of a \textit{specific} spin qubit within a quantum register. Such selective addressing using conventionally applied techniques, such as electron spin resonance,\cite{Koppens2006,Kroner2008} is rather challenging and can more conveniently be achieved with recently proposed electrical methods for spin control via Land$\rm\acute{e}$ g-tensor modulation.\cite{Pingenot2008,Andlauer2009,Pingenot2011} These approaches exploit tuning of the magnetic response by pushing the carrier envelope function into different regions of composition-engineered quantum dot nanostructures. Although electrical g-factor modulation has been successfully demonstrated in composition-graded AlGaAs quantum wells \cite{Salis2001} and vertically coupled InGaAs QD-molecules \cite{Doty2006} for several years, significant tuning of g-factors in individual self-assembled QDs was achieved only recently.\cite{Klotz2010,Jovanov2011} It was shown that a static electric field can be applied to quench the orbital angular momentum of the confined carriers, thus, modifying  the g-factor.\cite{pryor:2006,Jovanov2011,vanBree2011} Since most of the experiments addressing the spin of confined carriers are performed in magnetic fields it is important to develop a microscopic understanding of how \emph{magnetic fields} influence the quantum properties of the orbital states.

%----------------------------------------------------------------------------------------------------------------------------------------
In this paper, we report strong magnetic field induced tuning of the exciton g-factor in composition-engineered In$_{x}$Ga$_{1-x}$As-GaAs self-assembled QDs. By comparing our experimental results with realistic eight-band $\textbf{k}\cdot\textbf{p}$ simulations performed using QD size, shape and compositional information obtained from cross-sectional scanning tunneling microscopy (X-STM) we identify the origins of the magnetic field-dependence of the g-factor. Our results show that magnetic fields influence the excitonic g-factor via a mechanism that differs fundamentally from the case of static electric fields.\cite{Jovanov2011} In particular, the combination of the dilute In-composition ($x<0.35$) and comparatively large lateral size of the QDs leads to spin selective mixing of the lowest energy heavy hole (HH) and light hole (LH) orbital states, the strength of which is controlled by the external magnetic field. This gives rise to a quadratic Zeeman spin-splitting -- a phenomenon previously observed only in semiconductor quantum wells and superlattices.\cite{Snelling:1992, Warburton:1993}

%----------------------------------------------------------------------------------------------------------------------------------------
% DOTS AND SAMPLE DESIGN
%
%----------------------------------------------------------------------------------------------------------------------------------------
\begin{figure}[htbp]
\includegraphics[width=\columnwidth]{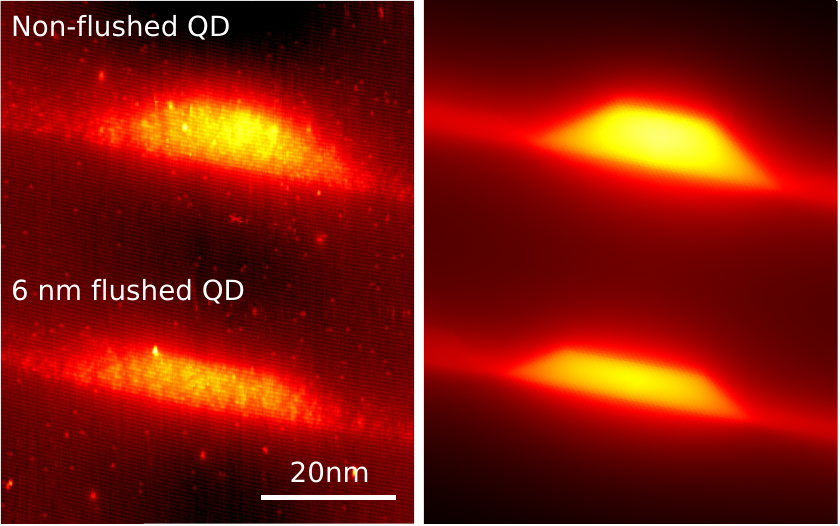}
\caption{\label{fig1} (color online). (left panel) Topography X-STM image of a conventionally grown quantum dot and a dot grown using identical conditions but including the In-flush technique. (right panel) Outward relaxation analysis of the two quantum dots from which the absolute In-concentration and spatial composition profile in the dots was extracted.}
\end{figure}
The samples investigated were electrically tunable GaAs \emph{n}-\emph{i}-Schottky photodiode structures into which two different types of QDs were embedded in the \emph{i}-region. Both samples were produced using molecular beam epitaxy under nominally identical growth conditions. The first sample was grown utilizing the conventional Stranski-Krastanov growth, whilst in the second was utilized the partially covered island (PCI) "In-flushing" method.\cite{Fafard1999} In the following, the sample containing the conventionally grown dots is referred to as \textit{non-flushed} and, analogously, the sample containing the dots grown with the PCI technique as \textit{flushed}. Both samples had a single layer of In$_{x}$Ga$_{1-x}$As self-assembled QDs grown in the $140$~nm thick $i$-region, with a relatively high growth temperature of $590^{~\circ}$C. The QD layer consisted of $8$~ML of In$_{x}$Ga$_{1-x}$As with a nominal In-content of $x=0.50$, deposited at a rate of $0.41$~ML/s and an As overpressure of $1.5\times10^{-5}$~mbar. The comparatively high growth temperature is expected to lead to an average In-content that is lower than the nominal $x=0.50$, due to the combined effects of In-desorption,\cite{Heyn2003a} interdiffusion with the GaAs matrix material and In segregation.\cite{Heyn2003b} Comparison of our results with the simulations provides strong support for this expectation, showing that the strong tunability of $g_{ex}$ is inextricably linked to a low average $x\approx0.35$ and the In-Ga alloy profile. For the \emph{flushed} sample a growth interruption was included after the QDs had been partially capped with a $6$~nm thick GaAs layer. During this growth interruption the temperature was increased to $650^{~\circ}$C and kept constant for $30$~s. After this, the temperature was again lowered to the nominal growth temperature and an additional capping layer of GaAs was deposited. X-STM measurements\cite{Keizer:2010} revealed that the QDs from the high density regions of the wafers exhibit inhomogeneous In-composition profiles with a relatively large cross sectional size of $40-50$~nm and a height of $4-8$~nm.
Typical topography X-STM image of two representative QDs is shown in the left panel of Fig.~\ref{fig1}.
%By simulating the outward relaxation of the cleaved facet we found that the In-concentration is increasing from the base to the apex.\cite{Bruls2002}
Two types of QD composition profiles were successfully fitted to the measured outward relaxation of the cleaved facet: (i) linearly increasing In-concentration from the base to the apex of the dot and (ii) an inverted trumpet like In-distribution.\cite{Bruls2002,offermans:2005, migliorato:2002} The results from the later are shown in the right panel of Fig.~\ref{fig1}. Here it should be noted that X-STM outward relaxation analysis can yield approximately similar concentration profiles that match to the surface relaxation and that the method itself does not provide an answer to which exact composition profile applies to the QDs studied.\cite{Mlinar2009} However, our experimental observations were found to be in good accord with theory only using the inverted trumpet In-distribution profile. The outward relaxation simulations revealed that for the conventionally grown \emph{non-flushed} QD shown in the left panel of Fig.~\ref{fig1}, the In-concentration in the apex is $x^{apex}=0.35$, reducing to $x^{min}=0.22$ at the base. In contrast, for the \emph{flushed} sample the In-concentration at the apex is slightly lower ($x^{apex}=0.30$) due to desorption of In during the flush step.

\begin{figure}[htbp]
\includegraphics[width=\columnwidth]{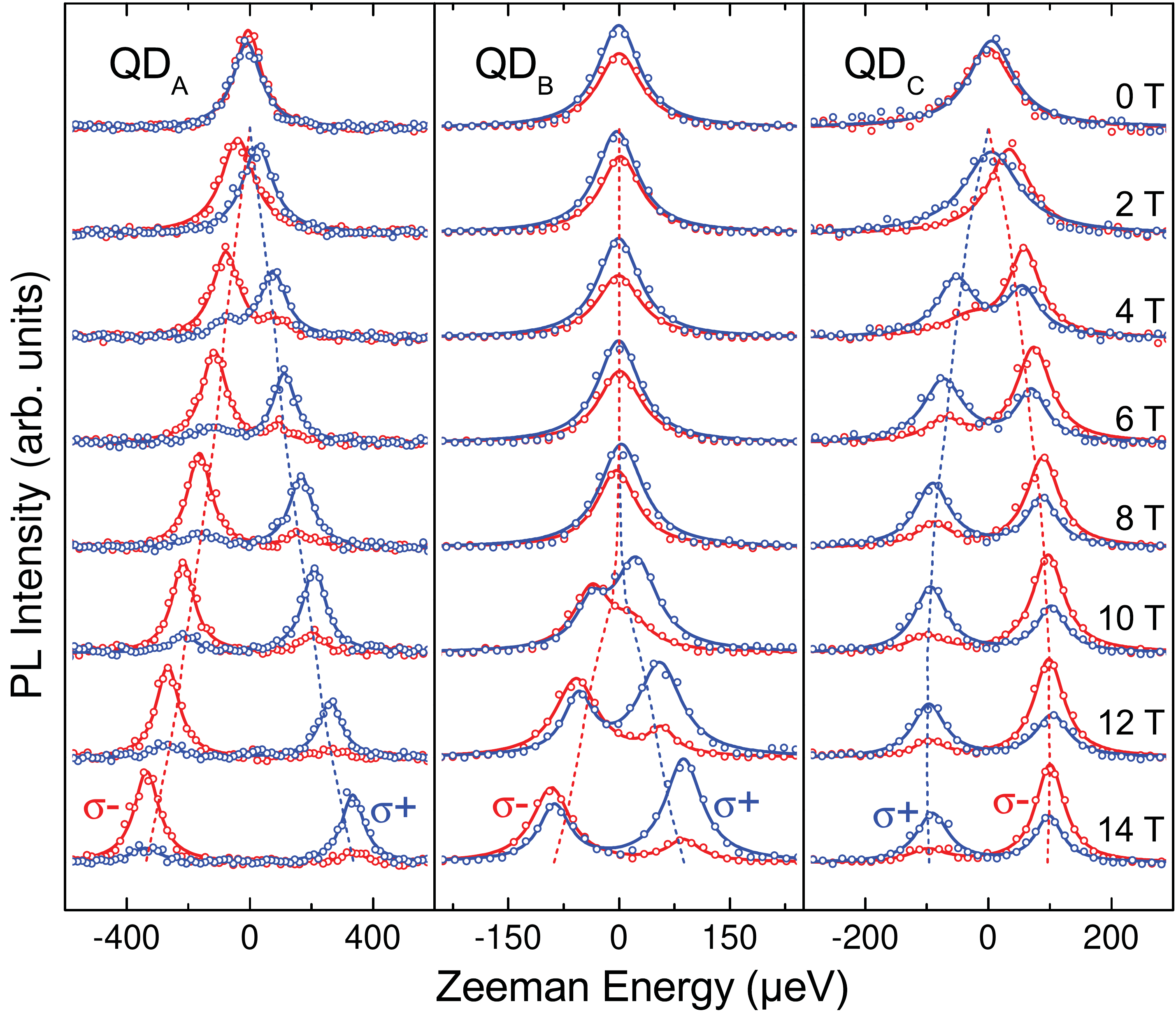}
\caption{\label{fig2} (color online). Polarization resolved photoluminescence spectra of three representative quantum dots grown without the partially covered island flushing technique.}
\end{figure}

\begin{figure}[htbp]
\includegraphics[width=\columnwidth]{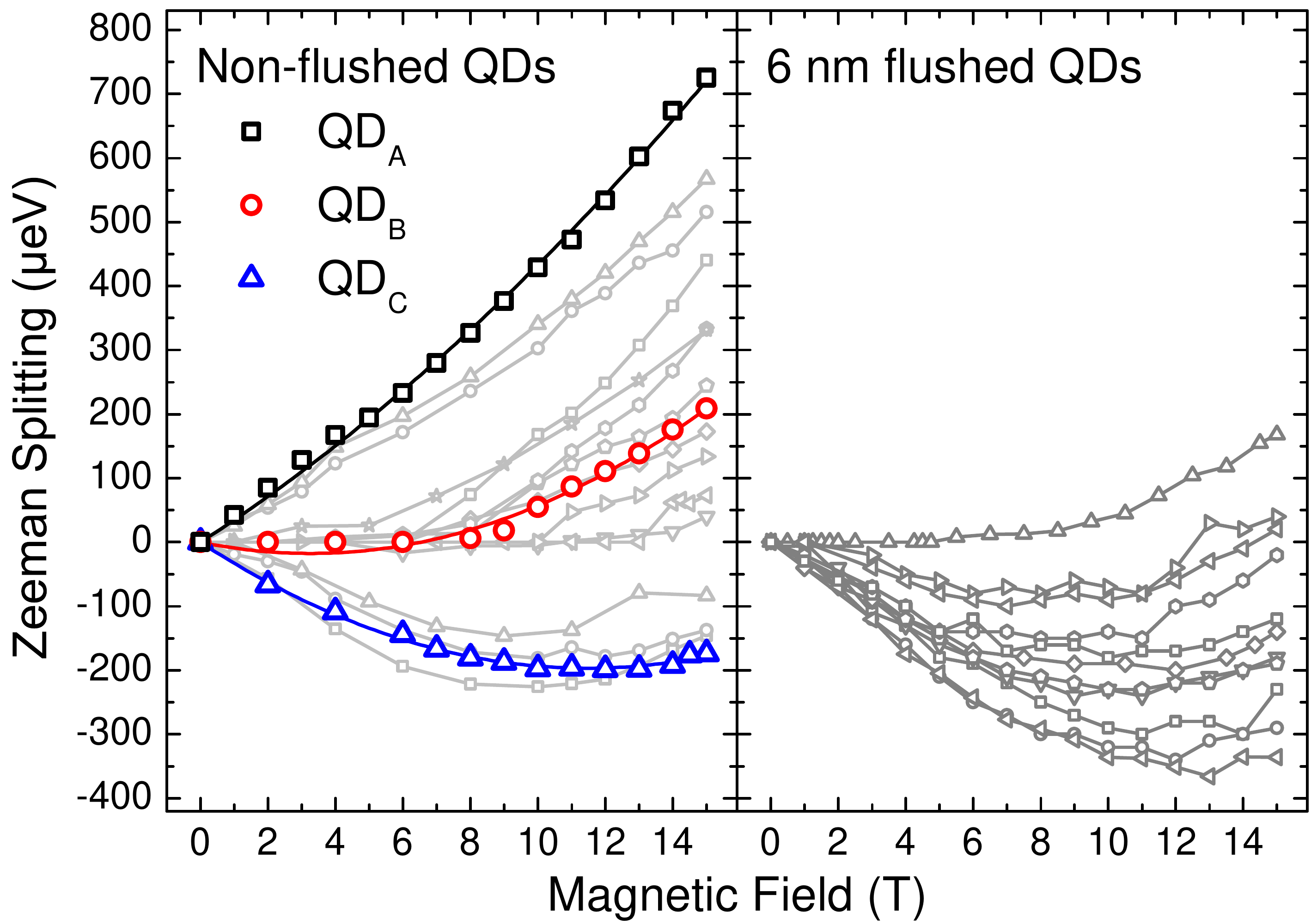}
\caption{\label{fig3} (color online). Zeeman splitting of the neutral exciton in several QDs grown without (left) and with the partially covered island flushing technique (right).}
\end{figure}

Optical characterization of the quantum dots was performed at low temperatures ($10$~K) using a confocal microscope. The microscope was placed in a superconducting magnet enabling application of magnetic fields up to $15$~T in Faraday configuration. Typical photoluminescence (PL) spectra are presented in Fig.~\ref{fig2} from the neutral exciton $X^{0}$ of three representative QDs from the \emph{non-flushed} sample. Many quantum dots ($>30$) were studied and the results summarized in Fig.~\ref{fig2} illustrate the full range of behaviors observed. The polarization-resolved PL spectra from the different QDs labeled as $\rm QD_{A}$, $\rm QD_{B}$ and $\rm QD_{C}$ reveal a substantially different behavior of the Zeeman splitting with increasing magnetic field. This can be clearly seen in the left panel of Fig.~\ref{fig3} where the Zeeman energy, defined as $\Delta E_{Z}=E(\sigma^{+}_{det})-E(\sigma^{-}_{det})=g_{ex}\mu_{B}B$, is plotted for the three dots, as well as for many other dots. Positive, negative as well as zero excitonic Zeeman splittings were observed for different dots from the same sample.
We show below that the observed variations reflect the range of size and material composition fluctuations measured in our X-STM microscopy studies. A striking feature in the Zeeman splittings presented in the left panel of Fig.~\ref{fig3} is the non-linear dependence of $\Delta E_{Z}$ on the magnetic field -- a magnetic \emph{field-dependent} exciton g-factor $g_{ex}=g_{e}+g_{h}=g_{ex}^{0}+g_{ex}^{1}B$. The best fit to the Zeeman splittings of $\rm QD_{A}$, $\rm QD_{B}$ and $\rm QD_{C}$ was obtained using quadratic function, as depicted by the solid lines presented in the left panel of Fig.~\ref{fig3}. The solid lines for the other QDs interpolate the experimental data. We attribute the pronounced non-linear dependence of $\Delta E_{Z}$ on B to a magnetic field-dependent hole g-factor $g_{h}$, as will be explained in detail below.

While the excitonic Zeeman splittings of the QDs from the \emph{non-flushed} sample revealed positive, negative and even zero B-field dependent excitonic g-factors, the dots investigated from the \emph{flushed} sample mainly revealed negative, but also B-field dependent g-factors. This is clearly demonstrated by the negative and non-linear Zeeman splittings presented in the right panel of Fig.~\ref{fig3}. The weaker variation in the Zeeman splittings observed for the dots from the \emph{flushed} sample compared to the \emph{non-flushed} sample arises from the reduced fluctuation in the dot height and the lower average In-concentration caused by the PCI growth process.

%----------------------------------------------------------------------------------------------
% THEORY
%
%----------------------------------------------------------------------------------------------

A weak non-linear dependence of the electron Zeeman splitting on the magnetic field was first observed for GaAs-AlGaAs quantum wells and superlattices at high magnetic fields.\cite{Dobers1988} Similar quadratic magnetic field dependencies of the hole Zeeman splitting have been found for GaAs-InGaAs superlattices and quantum wells and attributed to a magnetic field-induced mixing of HH and LH states.\cite{Warburton:1993, Traynor1995, Kotlyar:2001}
In quantum wells subject to magnetic field applied parallel to the quantization axis, the hole wave function can be factorized in a product of quantum well states in the vertical direction and Landau levels in the lateral direction.\cite{Pidgeon1966, Kotlyar:2001} A magnetic field applied along the $[001]$ direction induces a coupling of the $\rm HH\uparrow$ ground state with the $\rm LH\uparrow$ and the $\rm LH\downarrow$ bands with a strength that is proportional to $\gamma_3 \sqrt{B} \hat k_z$  and $(\gamma_2 + \gamma_3) B$, respectively.\cite{trebin:1979} The operator $\hat k_z$ acts in the growth direction and $\gamma_2$ and $\gamma_3$ are the Luttinger parameters.\cite{Luttinger1956} Interestingly, the $\rm HH\downarrow$ ground state couples only to the $\rm LH\uparrow$ band via an interaction that varies as $(\gamma_2 - \gamma_3) B$. Since $\gamma_2 \approx \gamma_3$ for In${_x}$Ga$_{1-x}$As alloys,\cite{Vurgaftman2001} the latter coupling is normally negligible and the $\rm HH\downarrow$ ground state consequently has an almost pure $HH\downarrow$ character, independent of the magnetic field.
It has been shown that the coupling of the $\rm HH\uparrow$ ground state to the light hole bands leads to a heavy-hole g-factor that varies with the square of the in-plane wave vector, i.e.  $g_h\propto k_\|^2$.\cite{Kotlyar:2001}
For quantum well Landau levels, $k_\|^2$ varies linearly with $B$ leading to the experimentally observed quadratic Zeeman splitting. On the other hand, in small, strongly confined quantum dots $k_\|^2$ varies with $\sim 1/D^2$ where $D$ is the dot diameter. As a result, the hole g-factor in strongly confined dots should, therefore, be unaffected by the magnetic field. Our results indicate that the large QDs with dilute In-composition produce effects that fall in a regime between the expectations for quantum wells ($g_{h}=g_{h}^{0}+g_{h}^{1}B$) and quantum dots ($g_{h}$ independent of B). %\textbf{XXX  (g_h=g_h^0+g^*B) and quantum dots (g_h independent of B) XXX}

To understand the microscopic origin of the observed non-linear Zeeman splitting we performed a detailed three dimensional electronic structure calculation using the eight-band $\textbf{k}\cdot\textbf{p}$ envelope function approximation.
In order to include the B-field in our calculations we used the recently proposed gauge invariant \textit{symmetry adapted finite element method} that accurately accounts for valence band couplings.\cite{Eissfeller:2011} Strain fields were included using continuum elasticity theory and their impact on the electronic structure was fully taken into account via deformation potentials and the linear piezoelectric effect.\cite{stier:1999} The exchange interaction is expected to be of minor importance for the neutral exciton because of the large effective band gap of $\approx1320$~meV and the weak mixing of conduction- and valence-bands. The direct Coulomb interaction was found to have a negligible influence on the exciton g-factor and is, therefore, also neglected in our simulations.\cite{Jovanov2011}

To obtain quantitative results for the $X^0$ g-factor, a Luttinger-like eight band $\textbf{k}\cdot\textbf{p}$-model was employed, where remote-band contributions to the effective mass Hamiltonian and g-factors are included up to the order $k^2$.\cite{trebin:1979} We modeled our QDs as having a truncated lens shape with a diameter varying from $D=25-50$~nm, a height of $4$~nm above the wetting layer (WL) and an inverse trumpet-like In-compositional profile.\footnote{$x({\rm x},{\rm y},{\rm z})=x^{min}+(x^{apex}-x^{min})exp[-\rho/\rho_{0}exp(-{\rm z}/{\rm z_{0}})]$, with $\rho=\sqrt{{\rm x}^{2}+{\rm y}^{2}}$, $\rho_{0}=0.3$~nm and ${\rm z_{0}}=1.5$~nm} The In-concentration of the In$_{x}$Ga$_{1-x}$As alloy was taken to be $x^{min}=0.2$ at the base and side of the dot increasing to $x^{apex}=0.3-0.5$ at the dot apex.\cite{migliorato:2002} These parameters are consistent with the results of X-STM measurements (Fig.~\ref{fig1}) performed on samples grown under the same conditions, from which we also determined the thickness and In-content of the wetting layer to be $2$~nm and $x^{WL}=0.18$, respectively.\cite{Keizer:2010}

\begin{figure}[htbp]
\includegraphics[width=\columnwidth]{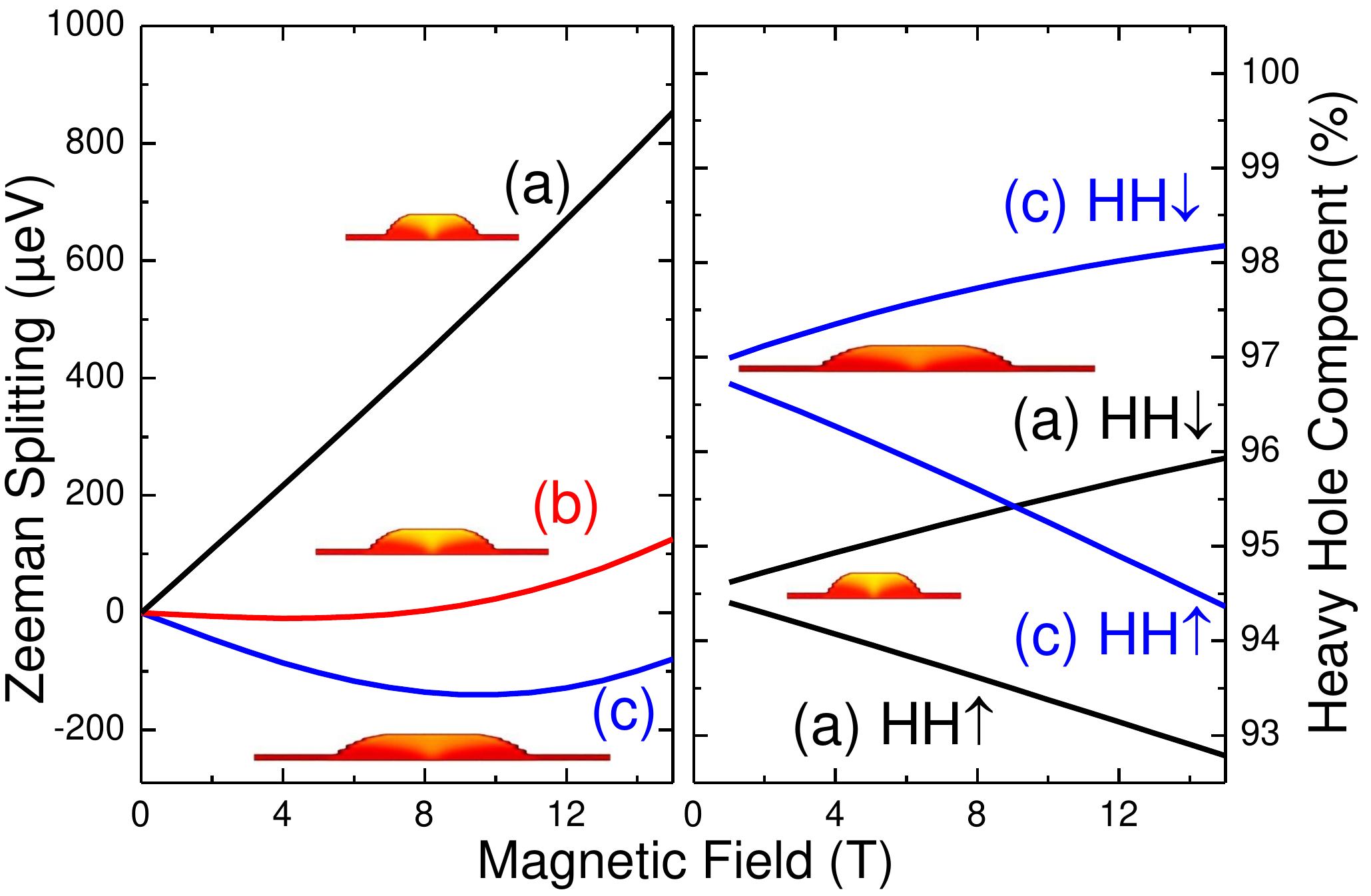}
\caption{\label{fig4} (color online). (left) Calculated Zeeman spin splitting of the neutral exciton in three QDs with the same height of $6~\textrm{nm}$ but different diameter and In-concentration at the apex $x^{apex}$ as a function of the magnetic field: (a) $25~\textrm{nm}$ diameter and $x^{apex}$=$0.50$, (b) $45~\textrm{nm}$ diameter and $x^{apex}$=$0.45$, (c) $50~\textrm{nm}$ diameter and $x^{apex}$=$0.40$. (right) Calculated heavy hole character of the lowest hole orbital levels having spin-up $\textrm{HH}\uparrow$ and spin-down $\textrm{HH}\downarrow$ character, respectively.}
\end{figure}

The left panel of Fig.~\ref{fig4} shows the calculated exciton Zeeman spin splitting as a function of the magnetic field for three model dots having different size and In-composition consistent with the range obtained from our X-STM measurements. These representative QDs have been chosen in order to reproduce the generic behavior of $\rm QD_{A}$, $\rm QD_{B}$, and $\rm QD_{C}$, representing the range of behaviors observed in our experiments. The In-concentration decreases from the data marked (a) to (c) in Fig.~\ref{fig4} from $x^{apex}$=$0.50$ to $x^{apex}$=$0.40$, whilst the lateral size increases from $25$~nm (a) to $50$~nm (c) to reproduce the experimentally observed exciton transition energies ($1310-1365$~meV). The curve labeled (a) in the left panel of Fig.~\ref{fig4} shows an almost purely linear Zeeman splitting (B-field independent g-factor) whilst (c) exhibits a clear quadratic dependence. The model QD (b) shows an intermediate behavior between the linear and quadratic regimes. The quadratic dependence of the exciton Zeeman splitting stems entirely from the HH-like lowest energy orbital state in the valence band.
For all QDs presented in Fig.~\ref{fig4} (a) to (c), the valence band Zeeman splitting varies quadratically with magnetic field, namely,
\begin{equation}
\Delta E_{Z}^{h} = \mu_B g_{h}^{0} B + \mu_B g_{h}^{1} B^2
\end{equation}
where the B-field is applied along the growth direction, $\mu_B$ is the Bohr magneton and, $g_{h}^{0}$ and $g_{h}^{1}$ are the linear and quadratic components of the hole g-factor, respectively.
%Consequently, a field-dependent hole g-factor can be defined as
%\begin{equation}
%g_h = g_{h,1} B + g_{h,0}.
%\end{equation}

These results indicate that the strong quadratic character of the hole Zeeman splitting ($g_{h}^{1}=0.016\pm0.002$~T$^{-1}$ for $\rm QD_{A}$, $0.028\pm0.002$~T$^{-1}$ for $\rm QD_{B}$ and $0.026\pm0.001$~T$^{-1}$ for $\rm QD_{C}$ from Fig.~\ref{fig3}) arises from the combination of the comparatively large diameter, small height and dilute In-content in the present dots. Firstly, the dot diameter is larger than the magnetic length over the entire range of B-fields of interest. Secondly, the low, almost homogeneous In-concentration induces only a weak confinement potential in the core of the QDs. As a result, the quantum states that are formed in magnetic field  resemble $2$D Landau levels and, consequently, the HH ground states behave in a manner similar to quantum wells. In addition, the small dot height of $6~\textrm{nm}$ (including the wetting layer) introduces a strong field-induced coupling of $\textrm{HH}\uparrow$-$\textrm{LH}\uparrow$ as in narrow quantum wells.\cite{Traynor1995} The quantum-well-like dependence of the Zeeman spin splitting on magnetic field is especially pronounced for the large, In-dilute QD (c) as shown in Fig.~\ref{fig4}. Due to the QD shape, In-Ga alloy profile and inhomogeneous strain fields, the lowest energy $\textrm{HH}$ orbital has a weak LH admixture at zero magnetic field. This is illustrated quantitatively in the right panel of Fig.~\ref{fig4}. However, as the B-field increases, the $\textrm{LH}$-admixture of the $\textrm{HH}\uparrow$-like ground state increases due to the field-induced LH-HH mixing alluded to above. In strong contrast, the $\textrm{HH}\downarrow$-like ground state effectively decouples from the $\textrm{LH}$-bands and, thus, its LH character weakens with the magnetic field.
%We find that the field induced mixing of the $\textrm{HH}\uparrow$-like ground state with $\textrm{LH}\uparrow$ is dominant confirming the qualitative picture of the system having almost quantum well character.
%However, we also observe a field induced mixing with the orbital state having $\textrm{LH}\downarrow$ in our calculations. This mixing is larger than expected from the higher order magnetic field coupling in the bulk semiconductor and we attribute it to the spatially varying alloy composition and strain field in the dot: As the B-field increases, the magnetic length decreases and the wave function becomes increasingly more strongly confined to the In-rich center of the dot changing the $LH\downarrow$ admixture.

%----------------------------------------------------------------------------------------------------------------------------------------
%----------------------------------------------------------------------------------------------------------------------------------------
% SUMMARY
%----------------------------------------------------------------------------------------------------------------------------------------
%----------------------------------------------------------------------------------------------------------------------------------------
In summary, strongly magnetic field-dependent exciton g-factors were observed in InGaAs self-assembled QDs. The microscopic origin of non-linear Zeeman splitting was accounted for by eight-band $\textbf{k}\cdot\textbf{p}$ simulations using realistic parameters (size and In-composition) that were directly extracted from X-STM measurements. The combined effect of dilute In-composition and relatively large dot lateral size was shown to result in strong field-induced mixing of the HH-LH orbital states in high magnetic fields. This mixing manifests itself as a quadratic variation of the hole Zeeman splitting on the external magnetic field. Similar effects are negligible for the electron and have previously been observed only in thin two dimensional systems.\\

This work is funded by the DFG via SFB-631, NIM, the TUM Institute for Advanced Study and the EU via SOLID.
\bibliography{g-fac_bib}

\end{document}